\documentclass[12pt,a4paper]{article}
\usepackage{latexsym,doublespace,psfig}
\usepackage{rawfonts}
\begin{document}
\newcommand{\npar}{\par\noindent}
\newcommand{\hef}{\ensuremath{^4He}}
\newcommand{\ahf}{\ensuremath{\overline{^4He}}}
\newcommand{\aht}{\ensuremath{\overline{^3He}}}
\newcommand{\beq}{\large\begin{equation}}
\newcommand{\eeq}{\end{equation}\normalsize}
\newcommand{\bma}{\large\[}
\newcommand{\ema}{\]\normalsize}
\newcommand{\kp}{{\textrm \ae}}
\newcommand{\Eq}[1]{Eq.\,(\ref{#1})}
\def\bear{\begin{eqnarray}}
\def\ear{\end{eqnarray}}
\newcommand{\F}{F_a}
\newcommand{\T}{\theta}
\newcommand{\sect}[1]{Sec.\,#1}
\newcommand{\Ref}[1]{Ref.\,\cite{#1}}
\newcommand{\Refs}[1]{Refs.\,\cite{#1}}

\pagestyle{empty}
\vspace*{2.0cm}

\centerline{\Large M.V.Lomonosov Moscow State University}
\centerline{\Large D.V.Skobeltsyn Institute of Nuclear Physics}

\vspace*{3.0cm}
\rightline{\large INP MSU Preprint 98--31/532}

\vspace*{2.0cm}
\centerline{\large K.M.~Belotsky{\small $^{2,3}$},
Yu.A.~Golubkov{\small $^{1,3}$},
M.Yu.~Khlopov{\small $^{2,3,4}$},}
\centerline{\large R.V.~Konoplich{\small $^{2,4}$}
and A.S.~Sakharov{\small $^{2,4}$}}

\vspace*{2.0cm}
\centerline{\LARGE Antihelium flux signature}
\centerline{\LARGE for antimatter globular 
cluster in our Galaxy}

\vspace*{1.0cm}
\centerline{\it {\small $^1$}Moscow State University, 
Institute of Nuclear Physics, Vorobjevy Gory,}
\centerline{\it 119899, Moscow, Russia,}
\centerline{\it {\small $^2$}Center for CosmoParticle Physics "Cosmion",}
\centerline{\it {\small $^3$}Institute of Applied Mathematics,
Miusskaya Pl.4, 125047, Moscow, Russia,}
\centerline{\it {\small $^4$}Moscow Engineering Physics Institute 
(Technical University), Kashirskoe Sh.31,}
\centerline{\it 115409, Moscow, Russia}

\vspace*{3.0cm}
\centerline{\large Moscow 1998}

\newpage
\vspace*{1.0cm}
\centerline{K.M.~Belotsky, Yu.A.~Golubkov{\small $^{a)}$},
M.Yu.~Khlopov{\small $^{b)}$},}
\centerline{R.V.~Konoplich{\small $^{c)}$}, 
A.S.~Sakharov{\small $^{d)}$}}

\bigskip
\noindent $^{a)}$e-mail: golubkov@elma01.npi.msu.su\\
\noindent $^{b)}$e-mail: mkhlopov@orc.ru\\
\noindent $^{c)}$e-mail: konoplic@orc.ru\\
\noindent $^{d)}$e-mail: sakhas@landau.ac.ru

\bigskip
\centerline{\large Preprint of Institute of Nuclear Physics 98---31/532}
\bigskip
\centerline{\Large Antihelium flux signature 
for antimatter globular cluster in our Galaxy}

\vspace*{1.0cm}
\begin{abstract}
	The AMS experiment is shown to be sensitive to test
the hypothesis on the existence of antimatter globular cluster 
in our Galaxy. 
The hypothesis follows from the analysis of possible tests 
for the mechanisms of baryosynthesis and uses antimatter domains 
in the matter dominated  Universe as the probe for the physics 
underlying the origin of the matter.
The interval of masses for the antimatter in our Galaxy is fixed 
from below by the condition of antimatter domain survival 
in the matter dominated Universe and from above by 
the observed gamma ray flux. For this interval the expected fluxes of 
antihelium-3 and antihelium-4 are calculated with the account for their 
interaction with the matter in the Galaxy.
\end{abstract}

\vspace*{6.0cm}
\leftline{\small{\copyright Moscow State University, 
Institute of Nuclear Physics, 1998}}

\newpage
\pagestyle{plain}
\bigskip
\leftline{\large{\bf 1. Antimatter in baryon asymmetrical Universe}}

The modern Big Bang theory is based on inflationary models 
with baryosynthesis and nonbaryonic 
dark matter. The physical basis for all three phenomena 
lies outside the experimentally proven theory of elementary particles. 
This basis follows from the extensions of the standard model.
Particle theory considers such extensions as aesthetical appealing
such as grand unification, as necessary to remove internal inconsistencies
in the standard model with the use of supersymmetry and axion or 
simply as theoretically possible ideas of 
neutrino mass or lepton and baryon number violation.
Most of these theoretical ideas can not be tested directly
and particle theory considers cosmological relevance as the
important component of their indirect test.
In the absence of direct methods of study one should analyse 
the set of indirect effects, 
which specify the models of particles and cosmology. 
The expected progress in the measurement of cosmic ray fluxes 
and gamma background and in the search for cosmic antinuclei 
makes cosmic ray experiments the important source of
information on the possible cosmological effects of particle theory. 
The first step in this direction
may be done on the base of AMS-Shuttle experiment. 
	
The specifics of AMS-Shuttle experimental programme 
puts stringent restriction on the possible choice of 
cosmic signatures for the new physics. At this stage
it can not be related to positrons, gamma rays or multi GeV
antiprotons. It
makes us to reduce the analysis to 
the antinuclear signal as the profound signature of new physics
and cosmology, related to existence of antimatter in the Universe. 

The generally accepted motivation for baryon asymmetric 
Universe is the observed absence of the macroscopic amounts of antimatter 
up to the scales of clusters of galaxies. According to the Big Bang theory
baryon symmetric homogeneous mixture of matter and antimatter 
can not survive after 
local annihilation, taking place at the first millisecond of 
cosmological evolution. Spatial separation of matter and antimatter 
can provide their survival in the baryon symmetric Universe
but should satisfy severe constraints on the effects of annihilation
at the border of domains. The most recent analysis finds that the
size of domains should be only few times smaller than the modern 
cosmological horizon to escape the contradictions with the observed
gamma ray background \cite{A}. 
In baryon asymmetric Universe the Big Bang theory predicts the 
exponentially small fraction of primordial
antimatter and practically excludes the existence of primordial
antinuclei. The secondary antiprotons may appear as a result 
of cosmic ray interaction with the matter.
In such interaction it is impossible to produce 
any sizeable amount of secondary antinuclei. 
Thus non exponentially small amount of antiprotons in the Universe 
in the period 
from $10^{-3}$ to $10^{16}$ s and antinuclei in the modern Universe
are the profound signature for new phenomena, related to 
the cosmological consequences of particle theory.

The inhomogeneity of baryon excess generation 
and antibaryon excess generation as the reflection of 
this inhomogeneity represents one of the most important example of such 
consequences.
It turned out \cite{1,2,3}, that practically all the existing 
mechanisms of baryogenesis can lead to 
generation of antibaryon excess in some places, when the baryon excess, 
averaged over the whole space, being positive.  So domains of 
antimatter in baryon asymmetric Universe provide a probe for the 
physical mechanism of the matter generation.

The original Sakharov's scenario of baryosynthesis \cite{4}
has found physical grounds in GUT models. It assumes CP violating effects
in out-of-equilibrium B-non-conserving processes, which generate 
baryon excess proportional to CP violating phase. 
If sign and magnitude of this phase varies  in 
space, the same out-of-equilibrium B-non-conserving processes, 
leading to baryon asymmetry, result in $B\,<\,0$ in the regions, where the 
phase is negative. The same argument is appropriate for the models of 
baryosynthesis, based on electroweak baryon charge 
nonconservation at high temperatures as well as 
on its combination with lepton number violation processes,
related to the physics of Majorana mass of neutrino. In all these
approaches to baryogenesis independent on the physical nature of 
B--nonconservation 
the inhomogeneity of baryon excess and generation of 
antibaryon excess is determined by the spatial dependence of CP 
violating phase. 

Spatial dependence of this phase is predicted in 
models of spontaneous CP violation, modified to escape the 
supermassive domain wall problem (see rev. in \cite{1,2} and Refs. therein).

In this type of models CP violating phase acquires discrete values 
$\phi_{+}=\phi_{0}+\phi_{sp}$ and $\phi_{-}=\phi_{0}-\phi_{sp}$, 
where $\phi_{0}$ and $\phi_{sp}$ are, respectively, constant and 
spontaneously broken CP phase, and antibaryon domains appear in the 
regions with $\phi_{-}<0$, provided that $\phi_{sp}>\phi_{0}$.  

In models, where CP violating phase is associated with the 
amplitude of invisible axion field, spatially-variable phase $\phi_{vr}$ 
changes continuously from $-\pi$ to $+\pi$. The amplitude of 
axion field plays the role of $\phi_{vr}$ in the period starting from 
Peccei-Quinn symmetry breaking phase transition until the axion mass 
is switched on at $T\,\approx\,1$ GeV. 
The net phase changes continuously and if 
baryosynthesis takes place in the considered period axion induced 
baryosynthesis implies continuous spatial variation of the baryon 
excess given by \cite{8}:

\beq
\label{A}
b(x)=A+b\sin\,\T (x).
\eeq

Here $A$ is the baryon excess induced by the constant CP-violating phase, 
which provides the global baryon asymmetry of the Universe and $b$ is 
the measure of axion induced asymmetry. If $b>A$, antibaryon excess 
is generated along the direction $\T=3\pi /2$. The stronger is the 
inequality $b>A$, the larger interval of $\T$ around the layer 
$\T=3\pi /2$ provides generation of antibaryon excess \cite{K}. 
In the case $b-A=\delta\ll A$ the antibaryon excess is proportional to 
$\delta^2$ and the relative volume occupied by it is proportional 
to $\delta$.  

The axion induced antibaryon excess forms the Brownian structure 
looking like an infinite ribbon along the infinite axion string 
(see \cite{6}). The minimal width of the ribbon is of the 
order of horizon in the period of baryosynthesis and is equal to 
$m_{Pl}/T^{2}_{BS}$ at 
$T\approx T_{BS}$. At $T<T_{BS}$ this size experiences red shift and 
is equal to 

\beq
\label{S}
l_h(T)\approx\frac{m_{Pl}}{T_{BS}T}\,.
\eeq

This structure is smoothed by the annihilation at the border of 
matter and antimatter domains. When the antibaryon diffusion scale 
exceeds $l_h(T)$ the infinite structure decays on separated domains. 
The distribution on domain sizes turns to be strongly model dependent 
and is calculated in \cite{7}. 

The size and amount of antimatter in domains, generated in the result
of local baryon-non-conserving out-of-equilibrium processes, 
is related to the parameters of models of CP violation and/or 
invisible axion (see rev. in \cite{1,3}).
SUSY GUT motivated mechanisms of baryon asymmetry imply flatness of 
superpotential relative to existence of squark condensate. Such a 
condensate, being formed with $B\,>\,0$, induces baryon asymmetry, after 
squarks decay on quarks and gluinos.  The mechanism doesn't 
fix the value and sign of B in the condensate, opening the 
possibilities for inhomogeneous baryon charge distribution and 
antibaryon domains \cite{3}. The size and amount of antimatter in
such domains is determined by the initial distribution of 
squark condensate.

So antimatter domains in baryon asymmetric Universe are related to 
practically all the mechanisms of baryosynthesis, and serve 
as the probe for the mechanisms of CP violation and primordial 
baryon charge inhomogeneity. The size of domains depends on the 
parameters of these mechanisms.

With the account for all possible mechanisms for 
inhomogeneous baryosynthesis, predicted on the base of
various and generally independent extensions of the standard
model, the general analysis of possible domain distributions
is rather complicated.
Fortunately, the test for the possibility of the existence 
of antistars in our Galaxy, offered in \cite{K}, 
turns to be practically model independent and as we 
show here may be accessible to cosmic ray experiments, to AMS
experiment, in particular.

\bigskip
\leftline{\large{\bf 2. Antimatter globular cluster in our Galaxy}}

Assume some distribution of antimatter domains,
which satisfies the constraints on antimatter annihilation
in the early Universe. Domains, surviving 
after such annihilation, should have the mass exceeding

\beq
\label{m}
M_{min}\approx (b/A)\rho_{b}l^3_a\,,
\eeq

where $\rho_b$ is the mean cosmological baryon density. The mass fraction
$f$ of 
such domains relative to total baryon mass is strongly model dependent.
Note 
that since the diffusion to the border of antimatter domain is determined
on RD 
stage by the radiation friction the surviving scale fixes the size of the 
surviving domain. On the other hand the constraints on the effects of 
annihilation put the upper limit on the mass of annihilated antimatter.

The modern antimatter domain distribution should be 
cut at masses given by 
the \Eq{m} due to annihilation of smaller domains and it is the 
general feature of any model of antibaryosynthesis in baryon 
asymmetrical Universe. The specific form of the domain distribution 
is model dependent. At the scales smaller than the \Eq{m} the 
spectrum should satisfy the constraints on the relative amount of 
annihilating antimatter. Provided that these constraints are 
satisfied one may consider the conditions for antimatter objects 
formation. One should take into account that the estimation of the 
annihilation scale after recombination (see \cite{7})  gives for this 
scale the value close to the Jeans mass in the neutral baryon gas after 
recombination. So the development of gravitational instability may 
take place in antimatter domains resulting in the formation of 
astronomical objects of antimatter.

Formation of antimatter object has the time scale being of the order 
of $t_f\approx (\pi G\rho)^{-1/2}$. The object is formed provided 
that this time scale is smaller than the time scale of its collision 
with the matter clouds. The latter is the smallest in the beginning 
of the object formation, when the clouds forming objects have large 
size. 

Note that the isolated domain can not form astronomical object 
smaller than globular cluster \cite{K}. The isolated anti-star can not be 
formed in matter surrounding since its formation implies the 
development of thermal instability, during which cold clouds are 
pressed by hot gas. Pressure of the hot matter gas on the antimatter 
cloud is accompanied by the annihilation of antimatter. 
Thus anti-stars can be 
formed in the antimatter surrounding only, what may take place when such 
surrounding has at least the scale of globular cluster.

One should expect to find antimatter objects among 
the oldest population of the 
Galaxy \cite{K}. It should be in the halo, since owing to strong
annihilation of 
antimatter and matter gas the formation of secondary antimatter objects in
the 
disc component of our Galaxy is impossible. So in the estimation of
antimatter 
effects we can use the data on the spherical component of our Galaxy as
well as 
the analogy with the properties of the old population stars in globular
clusters 
and elliptical galaxies.

In the spherical component of our Galaxy the antimatter globular cluster
should 
move with high velocity (what follows from the velocity dispersion in halo 
($v\approx 150$ km/s) through the matter gas with very low number density 
($n\approx 3\cdot 10^{-4}cm^{-3}$). Owing to small number density of
antimatter 
gas effects of annihilation with the matter gas within the antimatter
globular 
cluster are small. These effects, however, deserve special analysis for
future 
search for antimatter cluster as the gamma source. 

The integral effects of antimatter cluster may be estimated by the 
analysis of antimatter pollution of the Galaxy by the globular 
cluster of antistars.

There are two main sources of such pollution: the antistellar wind 
(the mass flow from antistars) and the antimatter Supernova 
explosions. The first source provides the stationary in-flow of 
antimatter particles with the velocity $10^7\div 10^8cm/s$ to the 
Galaxy. From the analogy with the elliptical galaxies, for which one 
has the mass loss $10^{-12}M_{\odot }$ per Solar mass per year, one can 
estimate the stationary admixture of antimatter gas in the Galaxy and 
the contribution of its annihilation into the gamma ray background. 
The estimation strongly depends on the distribution of magnetic 
fields in the Galaxy, trapping charged antiparticles. Crude 
estimation of the gamma flux from the annihilation of this antimatter 
flux is compatible with the observed gamma background for the total 
mass of antimatter cluster less than $10^5 M_{\odot }$. This estimation 
puts upper limit on the total mass fraction of antimatter clusters in our 
Galaxy. Their 
integral effect should not contradict the observed gamma ray background.

The uncertainty in the distribution of magnetic fields causes even more
problems 
in the reliable 
estimation of the expected flux of antinuclei in cosmic rays. It also is 
accomplished by the uncertainty 
in the mechanism of cosmic ray acceleration. The relative contribution of
disc 
and halo particles into 
the cosmic ray spectrum is also unknown. 

To have some feeling of the expected 
effect we may assume that the 
mechanisms of acceleration of matter and antimatter cosmic rays are similar
and 
that the contribution 
of antinuclei into the cosmic ray fluxes is proportional to the mass ratio
of 
globular cluster and Galaxy. Putting together the lower limit on the mass
of the 
antimatter globular cluster from the condition of survival of antimatter
domain 
and the upper limit on this mass following from the observed gamma ray 
background one obtains \cite{K}  the expected flux of antihelium nuclei in
the 
cosmic rays 
with the energy exceeding 0.5 Gev/nucleon to be $10^{-8}\div 10^{-6}$ of
helium 
nuclei observed in the cosmic rays. 

Such estimation assumes that annihilation does not influence the antinuclei
composition of cosmic rays, what may take place if the cosmic ray
antinuclei are 
initially relativistic. If the process of acceleration takes place outside
the 
antimatter globular cluster one should take into account the Coulomb
effects in 
the annihilation cross section of non relativistic antinuclei, what may
lead to
suppression of their expected flux.

On the other side antinuclei annihilation invokes new factor 
in the problem of their acceleration, which is evidently absent 
in the case of cosmic ray nuclei. 
This factor may play very important role in the account for antimatter 
Supernovae as the possible source of cosmic ray antinuclei. From the
analogy with elliptical galaxies one may expect \cite{K} 
that in the antimatter globular cluster Supernovae of the I type 
should explode with the frequency about 
$2\cdot 10^{-13}/{M_{\odot }}$ per year. 
On the base of theoretical models and observational 
data on SNI (see c.f. \cite{11}) one expects in such explosion the
expansion of a shell with the mass of about $1.4 M_{\odot }$ 
and velocity distribution up to $2\cdot 10^9cm/s$. 
The internal layers with the velocity $v<8\cdot 10^8cm/s$ 
contain anti--iron $^{56}Fe$ and the outer layers with higher velocity
contain lighter elements such as anti--calcium or anti-silicon. 
Another important 
property of Supernovae of the I type is the absence of hydrogen lines
in their spectra. Theoretically it is explained as the absence of
hydrogen mantle in Presupernova. In the case of antimatter Supernova
it may lead to strong relative enhancement of antinuclei relative
to antiprotons in the cosmic ray effect. Note that similar effect
is suppressed in the nuclear component of cosmic rays, since 
Supernovae of the II type are also related to the matter cosmic ray origin
in our Galaxy, in which massive hydrogen mantles (with the
mass up to few solar masses) are accelerated.  

In the contrast with the ordinary Supernova the expanding antimatter shell
is not decelerated owing to acquiring the interstellar matter gas and is not
stopped by its pressure but annihilate with it \cite{K}. 
In the result of annihilation with hydrogen, of which the matter gas 
is dominantly composed, semi--relativistic
antinuclei fragments are produced. The reliable analysis of such cascade of
antinuclei annihilation may be based on the theoretical models and
experimental data on antiproton nucleus interaction. 
This programme is now under way.
The important qualitative result is the possible nontrivial contribution 
into the fluxes of cosmic ray antinuclei with $Z\le 14$ and the enhancement of 
antihelium flux. With the account for this argument the estimation of
antihelium flux from its direct proportionality to the mass of antimatter
globular cluster seems to give the lower limit for the expected flux.

Here we study another important qualitative effect in the 
expected antinuclear composition of cosmic rays. Cosmic ray annihilation 
in galactic disc results in the significant fraction of 
anti-helium-3 so that antihelium-3 to antihelium-4 ratio turns to be the
signature of 
the antimatter globular cluster.

\bigskip
\leftline{\large{\bf 3. Equations for differential fluxes}}

Cosidering the $\ahf$ nuclei travelling 
through the Galactic disk we have to take into account two processes:

\begin{itemize}
\item[(i)] the destruction of a nucleus in the inelastic interactions
with the protons of the galactic media and
\item[(ii)] the energy losses during the travelling through the Galaxy.
\end{itemize}

For the $\aht$ nuclei we need to take into account also the possibility
of the $\aht$ nuclei production due to the reaction

\begin{itemize}
\item[(iii)]\large $\ahf\,+\,p\ \rightarrow\ \aht\,+\,all.$\normalsize
\end{itemize}

The energy losses occur due to four kinds of processes:

\begin{itemize}
\item[(a)] the energy losses on ionization and excitation of the
hydrogen atoms in the disk matter;
\item[(b)] the bremsstrahlung radiation on the galactic hydrogen atoms;
\item[(c)] the inverse Compton scattering on the relic photons and
\item[(d)] the synchrotron radiation in the galactic magnetic fields.
\end{itemize}

The processes (b) --- (d) are proportional to \large{$(m_e/M_{He})^2$}
and can be neglected at not very high energies of the $He$ nuclei.
The energy losses due to ionization and excitation of the hydrogen
atoms per one collision are being described by the expression
\cite{Landau4}:

\beq
\label{eloss}
\kp(\beta,z) \ = \ \frac{4\pi\,Z(z\alpha)^2}{m_e\,\beta^2}\,
\left [\ln\,\frac{2m_e\beta^2}{I\,(1-\beta^2)}\,-\,\beta^2\right ],
\eeq

\noindent where, $I$ is ionization potential of the hydrogen atom,
$I\,\approx\,15\ eV$; $Z\,=\,1,\ z\,=\,2$ are the electric charges
of the hydrogen and helium nuclei, respectively, $\beta\,=\,v/c$ is
the dimensionless velocity and
$\alpha\,=\,1/137$ is the fine structure constant.

\bigskip
The rates of the energy losses and the $\ahf$ nuclei destruction 
are:

\beq\label{rates}
\begin{array}{lll}
\frac{dE_{3,4}}{dt} & = & -n_H\,v_{3,4}\,\kp_{3,4}\\
\frac{dn_{3,4}}{dt} & = & -n_H\,v_{3,4}
\,\sigma_{ann}^{(3,4)}\,n_{3,4},
\end{array}
\eeq

\noindent where $n_H$ is the particle density of $H$ atoms in the
Galactic 
disc.

The source of $\aht$ nuclei can be written in the form:

\beq\label{he3in}
\frac{dn_3^{(+)}(t,E_3)}{dt} \ = \ -\,\int_{E_3}^{\infty}\,dn_4(t,E_4)
\,\frac{\partial W(E_4;E_3)}{\partial E_3}.
\eeq

\noindent $\partial W(E_4;E_3)/\partial E_3$ describes 
the probability to produce $\aht$ in the inelastic collision 
\hbox{$\ahf\,+\,p\,\rightarrow\,\aht\,+\,all$},
with the normalization condition:

\bma
\int_0^\infty\,dE_3\,\frac{\partial W(E_4;E_3)}{\partial E_3}\ =\ W_3(E_4).
\ema

If we introduce the differential flux 

\bma
J(t,E)\ =\ v\,\frac{\partial n(t,E)}{\partial E}
\ema

\noindent and the energy per nucleon ($E\,\rightarrow\,E/A$),
with $A\,=\,4$ --- the atomic weight of the anti--helium nucleus,
we obtain finally a system of the integro--differential equations,
describing the behaviour of $\ahf$ and $\aht$ nuclei in the Galaxy:

\beq\label{finsys}
\begin{array}{lll}
\frac{dJ(t,E_4)}{dt} & = & -n_H\,c\,\beta_4\,\left [\sigma_{inel}(p_4)\ 
-\ A\,\frac{m_p^2}{p_4\,E_4^2}
\,\frac{d\kp(\beta_4)}{d\beta_4}\right ]\,
J(t,E_4),\\
\\
\frac{dE_4}{dt} & = & -n_H\,c\,A^{-1}\,\beta_4\,\kp(\beta_4),\,\\
\\
\frac{dJ(t,E_3)}{dt} & = & -n_H\,c\,\beta_3
\,\left [\sigma_{inel}(p_3)\ 
-\ (A-1)\,\frac{m^2}{p_3\,E_3^2}
\,\frac{d\kp(\beta_3)}{d\beta_3}\right ]\,J(t,E_3)\\
\\
& &
+n_H\,c\,\beta_3\,{\displaystyle\int}_{E_3}^{\infty}\,dE_4\,\sigma_4(p_4)
\,\frac{\partial W(E_4;E_3)}{\partial E_3}\ J_4(t,E_4),\\
\\
\frac{dE_3}{dt} & = & -n_H\,c\,(A-1)^{-1}\beta_3\,\kp(\beta_3).
\end{array}
\eeq


\bigskip
\leftline{\large{\bf 4. The annihilation cross sections}}

Because the cross section of coherent interaction of the nucleon
with a nuclei is not larger than $(10\,-\,15)\%$ of the inelastic
cross section (see, e.g., \cite{Balestra93}), 
we can neglet such processes and put:

\beq\label{crsec}
\sigma_{ann}(N\,He)\ \approx\ \sigma_{inel}(N\,He),
\eeq

\noindent where, $\sigma_{ann}(N\,He)$ is the cross section for the
annilation of $\ahf$ at its collision with the nucleon and
$\sigma_{inel}(N\,He)$ is the inelastic cross section.

Total and elastic cross sections for the $pp$, $pn$, $\bar pp$, $\bar pn$
and $\bar pd$ ($d$ is the deutron) can be found in \cite{PDG96}.
For total cross sections at laboratory momentum $P_{lab}\,>\,50\ GeV/c$ 
we used the parametrization, following from the Regge fenomenology
\cite{PDG96}:

\beq\label{regge}
\sigma(p\,N)_{tot}\ =\ X\,s^{\epsilon}\ +\ Y\,s^{-\eta},
\eeq

\noindent where,

\vbox{
\beq
\begin{array}{lll}
X_{ab} & = & X_{\bar ab}\\
X_{pp} & = & 22.0\,\pm\,0.6\\
X_{pn} & = & 22.6\,\pm\,0.6\\
Y_{pp} & = & 56.1\,\pm\,4.4\\
Y_{\bar pp} & = & 98.2\,\pm\,9.5\\
Y_{pn} & = & 55.0\,\pm\,4.1\\
Y_{\bar pn} & = & 92.7\,\pm\,8.6\\
\eta & = & 0.46\,\pm\,0.3\\
\epsilon & = & 0.079\,\pm\,0.003\,.
\end{array}
\eeq
}

\normalsize
At $0.1\,<\,P_{lab}\,<\,50\ GeV/c$ we used plots from \cite{PDG96} 
for the total and elastic cross sections.

Very scare experimental data on total and elastic cross sections
for $p\,^4He$ can be found in 
\cite{Velichko85,Bujak81} and for $\bar p\,^4He$ 
in \cite{Balestra93,Balestra89,Balestra85}. Using these data
we found the $A$ dependence of the cross sections in the form:

\bigskip
\beq
\begin{array}{lll}
\sigma(^4He\,p) & = & A^{0.84}\times\,\frac{1}{2}
\,\left [\sigma(pp)+\sigma(np)\,\right ],\\
\sigma(\ahf\,p) & = & A^{0.84}\times\,\frac{1}{2}
\,\left [\sigma(\bar pp)+\sigma(\bar np)\,\right ].
\end{array}
\eeq

\begin{figure}[htb]                
\par
\centerline{\hbox{%
\psfig{figure=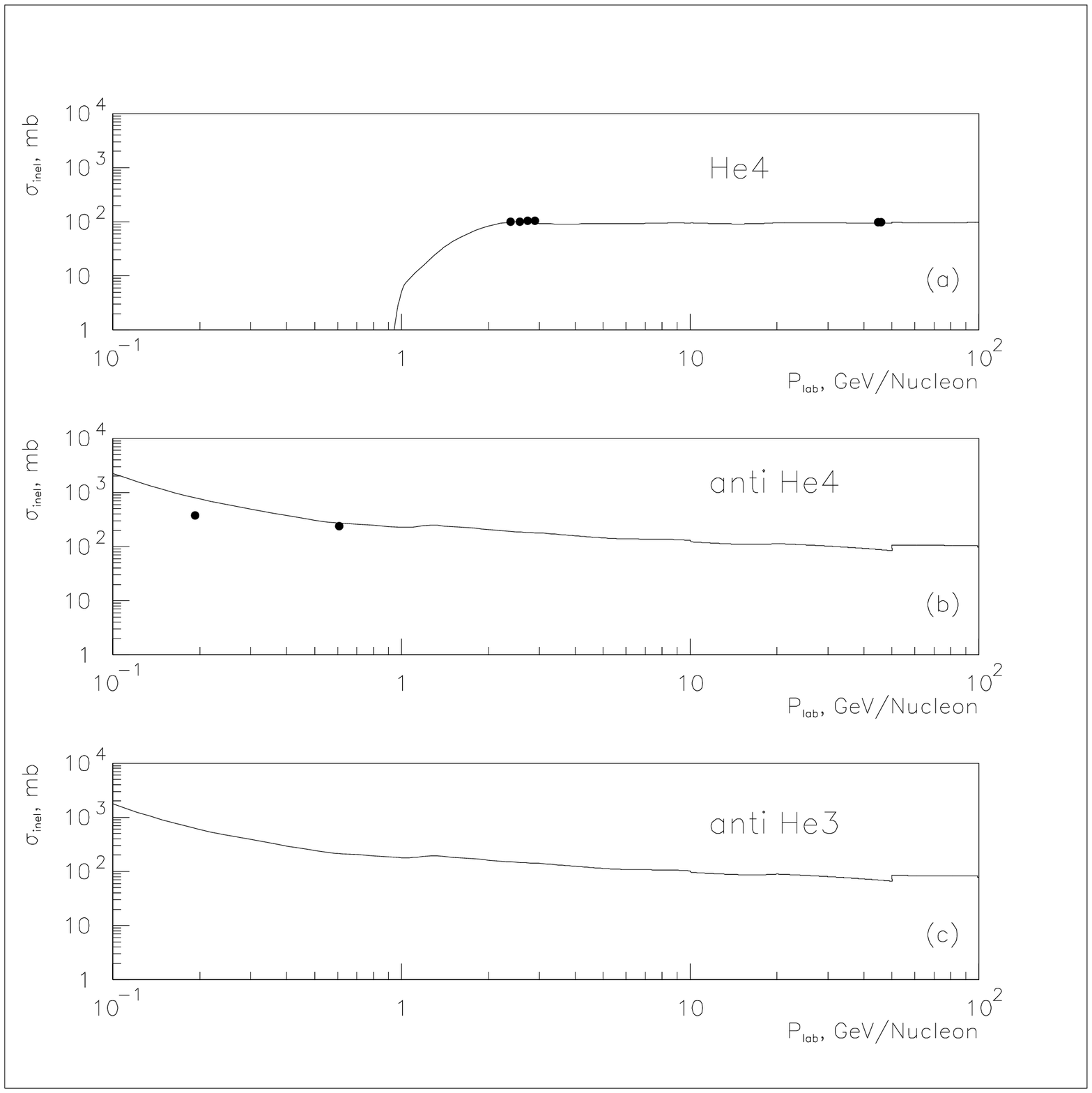,bbllx=1.5cm,bblly=5.5cm,%
bburx=19.0cm,bbury=23.0cm,clip=t,height=16.0cm}%
}}
\par
\caption{\label{crs}
Inelastic cross sections for: (a) ($^4He\ p$), (b) ($\ahf\ p$)
and  (c) ($\aht\ p$) interactions. The closed circles are the
experimental points for $\sigma_{inel}(p\,^4He)$ \cite{Velichko85,Bujak81}
and $\sigma_{inel}(\bar p\,^4He)$ \cite{Balestra93,Batusov90}.
}
\end{figure}

We used the above $A$--dependence also for the inelastic
cross section of $\aht\,p$ collisions.
The inelastic cross sections for interaction of $^4He$, $\ahf$
and $\aht$ with protons are shown in Fig.\ref{crs}. 
In this picture we also plotted the experimental
points for $\sigma_{tot}\,=\,\sigma_{el}$ of the
reactions $p\,^4He$ and $\bar p\,^4He$.


\bigskip
\leftline{\large{\bf 5. Results of the calculations}}

The experimental data from \cite{Balestra85,Balestra93} give
for the probability to produce the $\aht$ nucleus in $\ahf\,p$ collision:

\vbox{
\beq
\label{tohe3}
\frac{\sigma(\bar p\,^4He\,\rightarrow\,^3He\,+\,all)}
{\sigma_{ann}(\bar p\,^4He)}\ \approx
\ 0.25, \  at \ P\,=\,193\ MeV.
\eeq
}

We suggested that relative contribution
to $\aht$ does not depend on energy and used the above value.

For simplicity we suggested that the probability $dW(E_4;E_3)/dE_3$
in Eq.(\ref{he3in}) can be approximated by the \hbox{$\delta$--function}:

\bma
\frac{\partial W(E_4;E_3)}{\partial E_3}\ =\ W_3\ \delta(E_4\,-\,E_3),
\ema

\noindent with $W_3$ from Eq.(\ref{tohe3}).

\bigskip
The initial fluxes for $^4He$ and $\ahf$ we chose in the form:

\beq\label{influx}
\begin{array}{lll}
J_4(0,E) & = & 0.07\,\times\,\frac{1.93\,\beta}{E^{2.7}} \times 10^{-6},
\ cm^{-2}\,s^{-1}\,sr^{-1}\,\left (GeV/nucleon\right )^{-1},\\
\\
J_3(0,E) & = & 0.
\end{array}
\eeq

As the confinement time for $He$ nuclei in the galactic disc, where the
hydrogen number density is $n_H\,\approx\,1\ atom/cm^3$, we choose the
typical timescale $T_{conf}\ =\ 10^7\ yr$.
We also accounted for the very low density of the matter in the Galactic
halo.

\begin{figure}[htb]                
\par
\centerline{\hbox{%
\psfig{figure=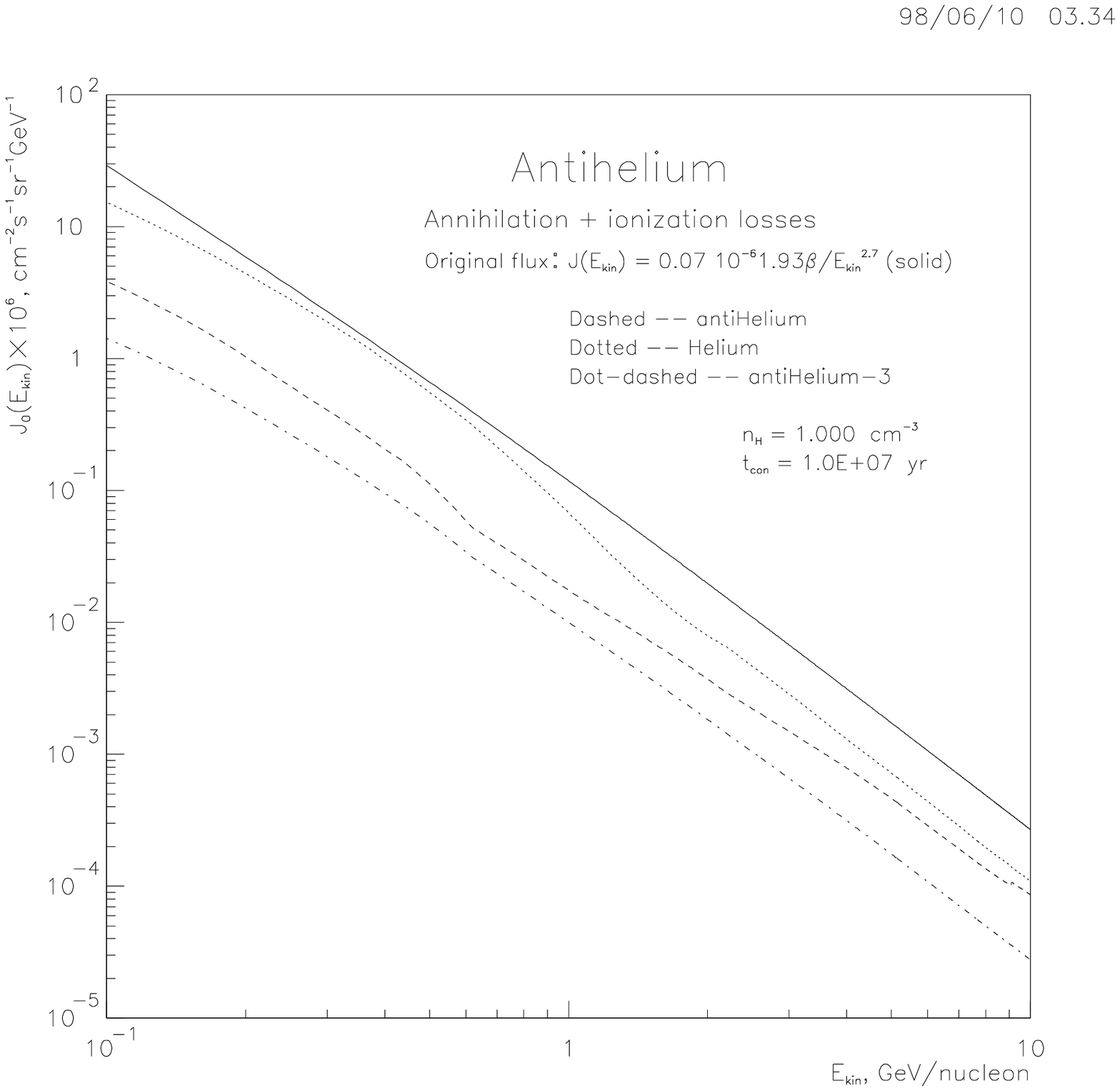,bbllx=1.5cm,bblly=5.5cm,%
bburx=19.0cm,bbury=23.0cm,clip=t,height=16.0cm}%
}}
\par
\caption{\label{fluxes}
Calculated fluxes of $\ahf$ (dashed), $^4He$ (dotted) and
$\aht$ (dash--dotted). Solid line presents initial flux for $^4He$
nucleii. The confinement time has been chosen equal to $10^7$ years.
}
\end{figure}

Results of our calculations are shown in Fig.\ref{fluxes}.
Solid line shows initial $He$ flux, dashed and dot--dashed lines represent
final fluxes of $\ahf$ and $\aht$, respectively. 

The first two equations in (\ref{finsys}) can be applied to the $\hef$
nuclei, 
if under the $\sigma_{ann}$ one understands the inelastic interaction
cross section of the $\hef$ nucleus with the proton again neglecting
the coherent processes.
For comparison we also
ploted by the dotted line the final flux of the $\hef$, suggesting
that the initial flux is the same as for $\ahf$.

\begin{figure}[htb]                
\par
\centerline{\hbox{%
\psfig{figure=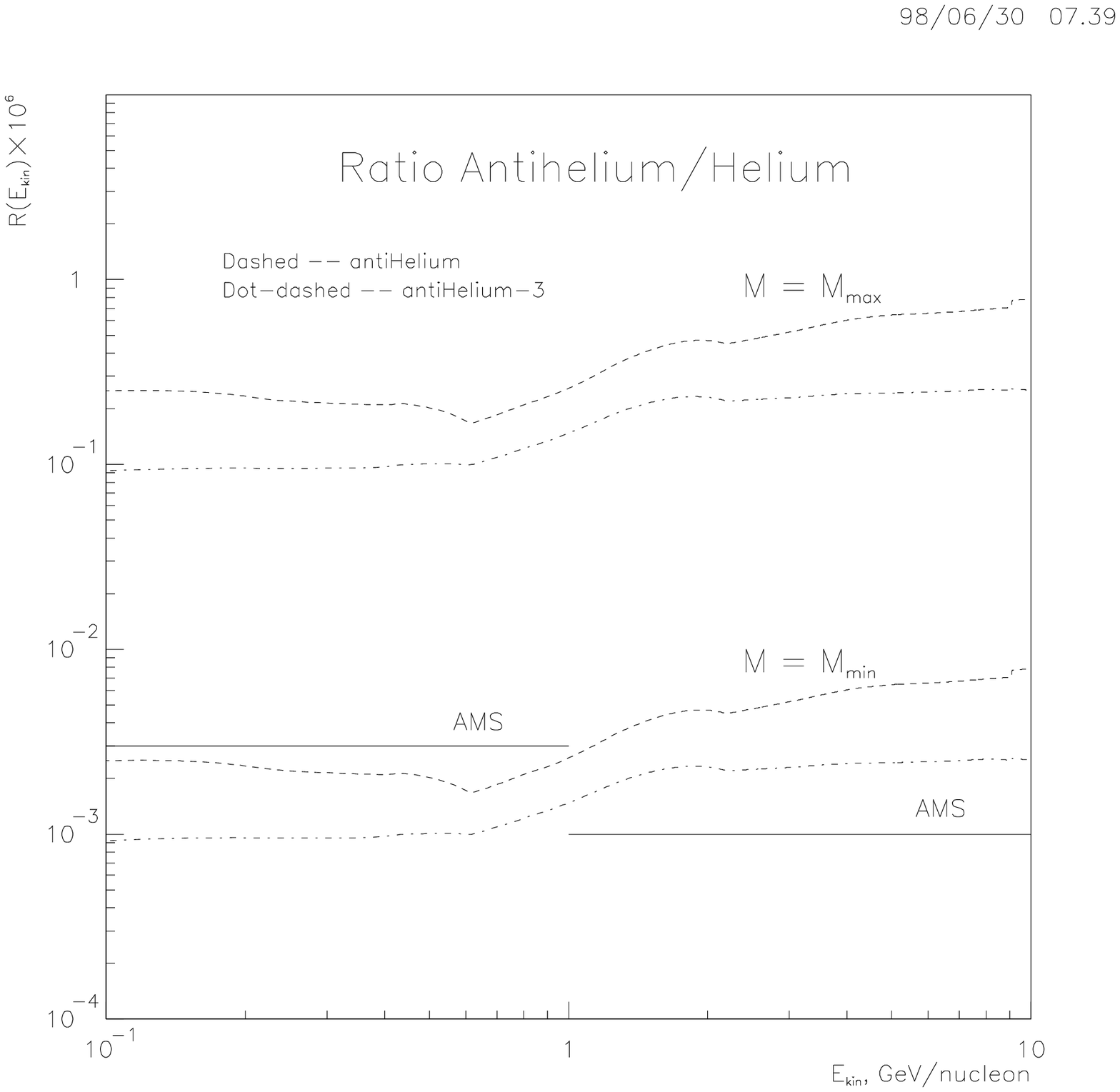,bbllx=1.5cm,bblly=5.5cm,%
bburx=19.0cm,bbury=23.0cm,clip=t,height=16.0cm}%
}}
\par
\caption{\label{ratios}
Ratios of fluxes $\ahf /^4He$ (dashed) and
$\aht /^4He$ (dash--dotted). Two upper curves correspond to the case of the
maximal possible mass of antimatter globular cluster 
$M_{max}\,=\,10^5\,M_{\odot }$ 
and the two lower curves to the case of the minimal possible mass 
of such cluster $M_{min}\,=\,10^3\,M_{\odot }$. 
The results of calculations are compared with the expected
sensitivity of AMS experiment \cite{Hoffer96} (solid lines).
}
\end{figure}

In Fig.\ref{ratios} we plotted the ratios of fluxes $\ahf /^4He$ and
$\aht /^4He$ for two cases: upper curves for $M/M_{MW}\,=\,10^{-6}$
and two lower curves for $M/M_{MW}\,=\,10^{-8}$. These results are compared
with the expected sensitivity of AMS experiment to antihelium flux. One
finds AMS experiment accessible to complete test of the hypothesis on the
existence of antimatter globular cluster in our Galaxy.


\bigskip
\leftline{\large{\bf 6. Discussion}}

The important result of the present work is that we found the substantial
contribution of antihelium-3 into the expected antinuclear flux. Even in
the case of negligible antihelium -3 flux originated in the halo its
contribution into the antinuclear flux in the galactic disc should be
comparable with the one of antihelium-4. 

The estimations of \cite{K}, on which our calculations are based,  assumed
stationary in-flow of antimatter in the 
cosmic rays. In case Supernovae play the dominant role 
in the cosmic ray origin the in--flow is defined by their frequency. 
One may find from \cite{K}  that the 
interval of possible masses of antimatter cluster 
$3\cdot 10^{3}\div 10^5 M_{\odot }$ gives 
the time scale of antimatter in-flow 
$1.6\cdot 10^{9}\div 5\cdot 10^{7}$ years, 
which exceeds the generally estimated life time of cosmic rays in the
Galaxy.
The succession of antinuclear annihilations may result in this case in the 
dominant contribution of antihelium and in particular antihelium-3 into the
expected antinuclear flux. It makes antihelium signature sufficiently
reliable even in this case.

Thus with all the reservations mentioned above on the base of the 
hypothesis on antimatter globular cluster in our Galaxy 
one may predict at the level of the expected 600 antiprotons 
up to ten antihelium events in the AMS--Shuttle experiment. 
Their detection would be exciting indication favouring
this hypothesis. Even the upper limit on antihelium flux will lead to important 
constraint on the fundamental parameters of particle theory and cosmology
to be discussed in our successive publications. 

Note that the important source of background for antinuclear events in
AMS-Shuttle experiment may be cosmic antiproton interaction with the matter
of the shuttle. Such interaction should give significant back-directed flux
of helium-4 imitating antihelium events in AMS detector. 
To have a feeling of this effect we may use the results of
the numerical simulations by Lozhkin and Kramarovsky \cite{Lozhkin98}
estimated the secondary nuclei multiplicities in the antiproton--iron
interactions.
According to these estimations which can be qualitatively correct at least 
for not very heavy nuclei the He-3 to He-4 ratio
in such interactions does not exceed 1:8. 
Moreover on the
contratry to the case of antinuclei back-directed nuclear flux contains
significant admixture of metastable isotopes, tritium, in particular.
According to Lozhkin-Kramarovsky calculations tritium to helium-4 ratio
reaches in this case 1:3.5, what may be important for the removal of
background events from the experimental data.
Another interesting feature of the secondary nuclei multiplicity
distributions is that being peaked at $z\,=\,2$, it exceeds the level
of $5\%$ for $z\,\le\,6$ and then falls down to $(1-2)\%$ for higher $z$'s,
giving negligible output for $z\,>\,18$.

We express our gratitude to H.Hofer, Ya.M.Kramarovsky, O.M.Lozhkin,
S.G.Rubin, A.L.Sudarikov and J.Ulbricht for fruitful discussions 
and to ETHZ for
the permanent support of studies undertaken on the base of Cosmion-ETHZ
collaboration. The work was performed in the framework  
of International projects "Astrodamus", Eurocos-AMS and Cosmion-ETHZ.

\vfill\eject


\vspace*{2cm}

\vspace*{2cm}

\vspace*{2cm}

\end{document}